\documentclass{article}
\usepackage{spconf,amsmath,graphicx}

\usepackage{enumitem}
\setlist{nosep, leftmargin=14pt}
\usepackage{xcolor}
\usepackage{mwe} 
\usepackage{graphics}      
\usepackage{booktabs}

\usepackage{todonotes}

\usepackage{lscape} 
\usepackage{siunitx,tabularx,ragged2e,booktabs,caption}
\usepackage{adjustbox}
\usepackage{lscape}

\makeatletter
\newcommand{\fmtnum}[1]{\if\relax\detokenize{#1}\relax\else\@fmtnum#1\relax\fi}
\def\@fmtnum#1.#2\relax{\eqmakebox[int][r]{$#1.$}\eqmakebox[dec][l]{$#2$}}
\makeatother
\usepackage{subfig}

\title{Addressing Deep Learning Model Calibration Using Evidential Neural Networks and Uncertainty-Aware Training}
\name{Tareen Dawood , Emily Chan, Reza Razavi, Andrew P. King$^{\star}$, Esther Puyol-Ant\'on$^{\star}$\thanks{*Joint last authors.}}
\address{School of Biomedical Engineering \& Imaging Sciences, King\textquotesingle s College London, UK}
\begin{document}
\maketitle
\begin{abstract}
In terms of accuracy, deep learning (DL) models have had considerable success in classification problems for medical imaging applications. However, it is well-known that the outputs of such models, which typically utilise the SoftMax function in the final classification layer can be over-confident, i.e. they are poorly \emph{calibrated}. Two competing solutions to this problem have been proposed: uncertainty-aware training and evidential neural networks (ENNs). In this paper we perform an investigation into the improvements to model calibration that can be achieved by each of these approaches individually, and their combination. We perform experiments on two classification tasks: a simpler MNIST digit classification task and a more complex and realistic medical imaging artefact detection task using Phase Contrast Cardiac Magnetic Resonance images. The experimental results demonstrate that model calibration can suffer when the task becomes challenging enough to require a higher capacity model. However, in our complex artefact detection task we saw an improvement in calibration for both a low and higher capacity model when implementing both the ENN and uncertainty-aware training together, indicating that this approach can offer a promising way to improve calibration in such settings. The findings highlight the potential use of these approaches to improve model calibration in a complex application, which would in turn improve clinician trust in DL models.
\end{abstract}
\section{Introduction}
\label{sec:intro}

In recent years, artificial intelligence (AI), and more specifically deep learning (DL) models have dominated medical imaging research for various classification problems \cite{chen2022recent}. In many such applications, DL will likely be utilised as a decision support tool to enable clinical users to make informed decisions with AI assistance that comply with evidence-based medical practice \cite{jin2022guidelines}. In this context, as well as the model prediction, the confidence of this prediction becomes a critical component of the AI model.

The concept of \emph{model calibration} refers to the relationship between the accuracy of predictions and their confidence: a well-calibrated model will be less confident when making wrong predictions and more confident when making correct predictions. With this in mind, measures of model calibration have been proposed that provide a more complete understanding of the performance of a model by estimating how closely the confidence matches the accuracy \cite{nixon2019measuring}. 

Good model calibration is of particular importance in medical imaging-based AI. Patient prognosis may be adversely affected if poorly calibrated models are deployed in a clinical setting, e.g. if models produce confident predictions for incorrect diagnoses. Therefore it is important that AI models for medical imaging are evaluated with appropriate metrics that can quantify calibration to more suitably assess a model's performance. Well-calibrated models for medical AI applications can provide an opportunity to develop safe and more reliable clinical decision-support tools \cite{reinke2023understanding}.

It is well-known that DL models which utilise the SoftMax function in their final layer can be poorly calibrated \cite{kompa2021second, gawlikowski2021survey}.
In the literature, two types of approaches have been proposed to attempt to improve the calibration performance of DL models: evidential neural networks (ENN) and uncertainty-aware training. ENNs \cite{sensoy2018evidential} are deterministic models that remove the dependency on the point estimates of the predictive distribution for confidence or uncertainty when using the SoftMax function. ENNs utilise probability distributions to obtain uncertainty in predictions for each category/class by gathering evidence during training using a Dirichlet distribution and have shown promising results towards improving model calibration in computer vision tasks \cite{guo2017calibration, bao2021evidential, minary2017face, tomani2021towards, karandikar2021soft}.

Uncertainty-aware training involves modifying the training of a standard DL model (e.g. by changing the loss function) to explicitly encourage weight updates that improve calibration. For example, \cite{krishnan2020improving} proposed an additional loss term based on the relationship between accuracy and uncertainty. \cite{bao2021evidential} utilised a similar approach but tailored their loss to be more suitable for their particular computer vision application. In the medical imaging domain interest in uncertainty-aware training has also grown with recent work \cite{dawood2021uncertainty} proposing a technique for improving calibration of a model for predicting response to treatment from cardiac magnetic resonance images. In medical image segmentation a novel uncertainty-aware framework was developed with a student and teacher model, where the student model learns from the teacher model by minimizing a segmentation loss and a consistency loss \cite{yu2019uncertainty}.

\textbf{Contributions:} In this paper, for the first time, we investigate and compare the use of ENNs and uncertainty-aware training (as well as their combination) to improve model calibration, and furthermore evaluate performance using both a low capacity model (LeNet5) and a higher capacity model (ResNet18). We perform two experiments to test our approaches: (1) a simple application in which we aim to classify handwritten digits using the publicly available MNIST database; and (2) a more complex and realistic medical application, in which we aim to detect the presence of artefacts in 2D flow Phase Contrast Cardiac Magnetic Resonance (PC-CMR) cine images.

\section{Materials and Methods}
\label{MnM}

Two different experiments were performed to evaluate the impact of different methods to address the problem of model calibration. 
For the first experiment, we used the publicly available MNIST digit dataset, which contains 50,000 training. 10,000 validation and 10,000 test images. For the second experiment, we utilised two retrospective 2D flow PC-CMR databases, in which all images were manually annotated for the presence of artefacts. The first database contained images from 80 patients from the UK Biobank \cite{sudlow2015uk} and the second database was retrieved from the clinical imaging database of Guy’s and St Thomas’ NHS Foundation Trust (GSTFT) and contained images from 830 patients with a range of different cardiovascular diseases acquired under routine clinical PC-CMR practice. For the second experiment, we combined the two databases and used 80\% of the total data for training, 10\% for validation, and 10\% for testing. Further details about the data and protocol can be found in \cite{https://doi.org/10.48550/arxiv.2209.14212}.

We evaluated the performance of two different DL models: a low-capacity model (LeNet5 \cite{lecun1998gradient}) and a higher capacity model (ResNet18 \cite{he2016deep}). To address model calibration, we utilised three approaches. The first approach was an ENN \cite{sensoy2018evidential}. The second approach was the \emph{Uncertainty versus Accuracy} (UvAC) uncertainty-aware training approach, which uses a differentiable loss function that quantifies the relationship between accuracy and uncertainty \cite{krishnan2020improving}. This loss function aims to improve uncertainty calibration and can be used as an additional penalty term in model training. The final approach was a combination of the ENN and UvAC. The aim here was to determine if the ENN's more robust estimate of uncertainty would improve the performance of the uncertainty-aware training method.

\section{Experiments and Results}
\label{Experiments}
We used the test set accuracy to measure model performance. To quantify model calibration we utilised the Adaptive Expected Calibration Error (AECE) metric \cite{ding2020revisiting}, again computed on the test set.\\

\noindent
\textbf{Experiment 1: MNIST} \\
In this experiment our motivation was to systematically modify model performance and calibration by adding noise of varying levels, and then to observe the impact of different methods aimed at improving calibration. After normalising the image intensities to a mean of zero and standard deviation of one, we added Gaussian noise (with mean of zero and different standard deviations) to both the train and test sets, and for each noise level the test accuracy and AECE were computed for each evaluated model. The values of the learning rate, number of epochs and batch size were taken from typical values in the literature: $\mathrm{10^{-3}}$, $10$ and $128$ (research has shown that the MNIST dataset is easily trainable with low error rates seen at 3-6 epochs \cite{ciresan2011flexible}). The AvUC loss weight and ENN scaling factor were optimised using a grid search strategy (ranges 0-4 and 0-40 respectively). See Table \ref{table:MNIST_Flow_Hyper_params} (left) for final hyperparameter values, which were selected to maximise validation set accuracy.
Table \ref{Evaluation Results MNIST Flow} (left) shows the results from this experiment, which are averaged over 10 train/test runs for each approach.\\

\begin{table}[htb]
  \caption{Summary of hyperparameters for Experiment 1 (MNIST) and Experiment 2 (PC-CMR) for all approaches using the LeNet and ResNet18 architectures.}
  \label{Evaluation Results MNIST Flow}
  \centering
  \small
  \resizebox{0.48\textwidth}{!}{\begin{tabular}{l||c|c||c|c}
    \toprule
    \midrule
    &\multicolumn{2}{c||}{\textbf{MNIST}} 
    &\multicolumn{2}{c}{\textbf{PC-CMR}}\\
    \textbf{Method} 
     & \textbf{AvUC} & \textbf{ENN} & \textbf{AvUC} & \textbf{ENN} \\
        \midrule
        \textbf{LeNet5+UvAC} & 2 & -  & 1.5 & -\\ 
        \textbf{LeNet5+ENN}  & - & 30  & - & 10\\ 
        \textbf{LeNet5+ENN+UvAC} & 1 & 40 & 2 & 10\\ 
        \midrule
        \textbf{ResNet18+UvAC} & 2 & - & 2 & -\\  
        \textbf{ResNet18+ENN} & - & 30 & - & 10\\  
        \textbf{ResNet18+ENN+UvAC} & 2 & 40 & 1 & 10\\ 
        \bottomrule
    \end{tabular}}
\label{table:MNIST_Flow_Hyper_params}
\end{table}

\begin{figure}
\vspace{-3mm}
    \centering
    \includegraphics[width=.45\textwidth, height=0.26\textwidth]{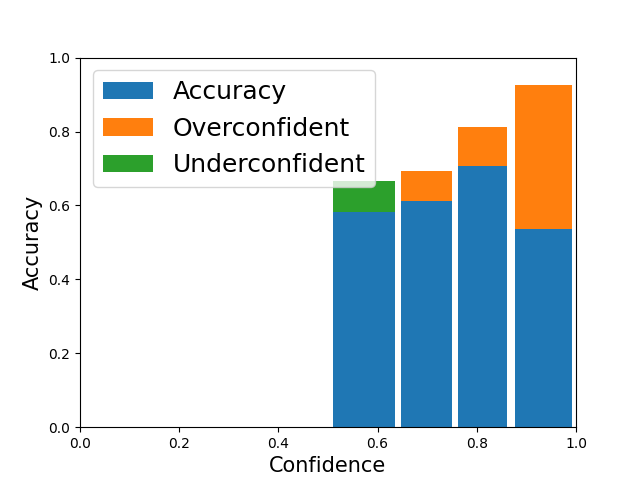}
    \centerline{(a) ResNet18}
    \includegraphics[width=.45\textwidth, height=0.26\textwidth]{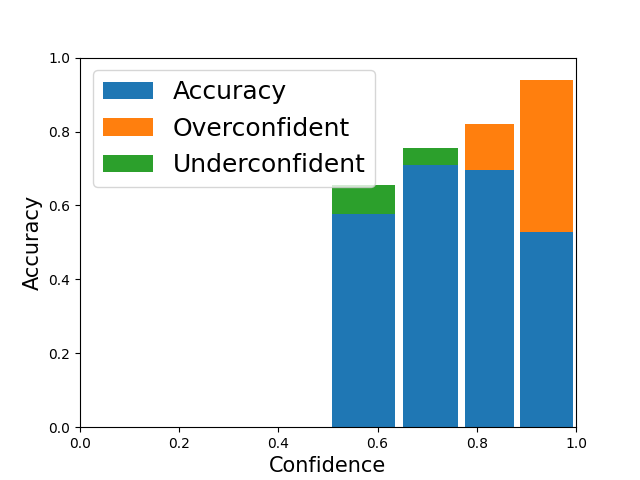}
    \centerline{(b) ResNet18+UvAC}
    \includegraphics[width=.45\textwidth, height=0.26\textwidth]{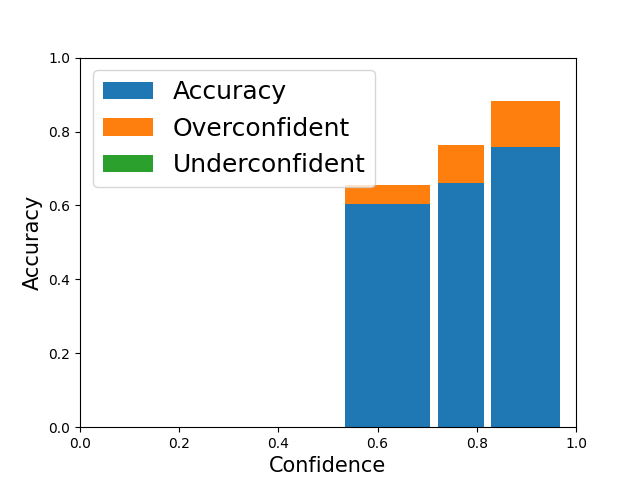}
    \centerline{(c) ResNet18+ENN}
    \includegraphics[width=.45\textwidth, height=0.26\textwidth]{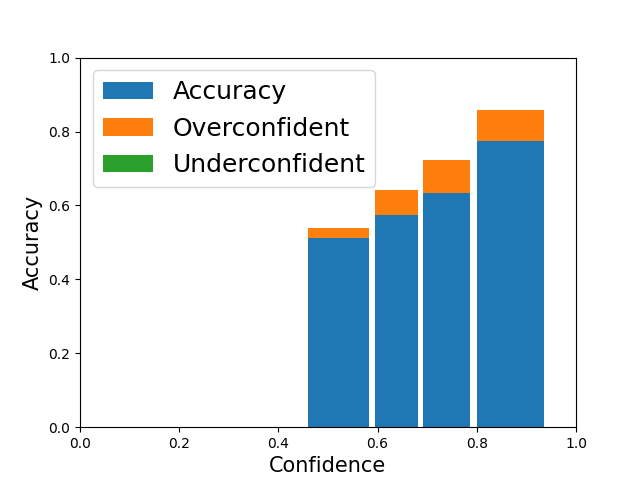}
    \centerline{(d) ResNet18+ENN+UvAC}
    \caption{AECE reliability diagrams for PC-CMR data with observed improvement wrt AECE when training with (a) baseline ResNet18 model versus (b) ResNet18 with the UvAC loss only (c) ResNet18 with ENN only and lastly (d) ResNet18 with ENN and UvAC loss.}
\label{fig:f1}
\end{figure}

\noindent
\textbf{Experiment 2: Phase Contrast Cardiac Magnetic Resonance (PC-CMR) Images} \\
Here we utilised a realistic clinical data set to address a more complex problem and observe the outcomes from our approaches. The problem of artefact detection from PC-CMR was chosen because these images are utilised by clinicians to quantify and assess cardiac function, but they can contain artefacts which make their analysis unreliable. Identifying these artefacts using DL models is a challenging problem.
Similar to Experiment 1 we quantified performance using the test accuracy and AECE. The values of the learning rate, number of epochs and batch size ($\mathrm{10^{-3}}$, $200$ and $16$ respectively) were taken from from values used in \cite{https://doi.org/10.48550/arxiv.2209.14212}, whilst the AvUC loss weight and ENN scaling factor were optimised using a grid search strategy (ranges 0-4 and 0-40 respectively). See Table \ref{table:MNIST_Flow_Hyper_params} (right) for final hyperparameters, which were selected to maximise validation set accuracy. Table \ref{table: Evaluation Results MNIST PC CMR Flow} (right) shows the results from this experiment, which were again averaged over 10 train/test runs for each approach. Sample reliability diagrams using AECE are shown in Figure \ref{fig:f1}, which illustrate the relationship between confidence and accuracy for all four approaches for the artefact classification problem.

\section{Discussion and Conclusion}
\label{DC}

We observe that there is inconsistency of performance across the two experiments. On the MNIST experiment, both the LeNet5 and ResNet18 models show high accuracy, but the ResNet models were more robust to noise. Both the LeNet5 and ResNet18 baseline models were well calibrated, and applying the ENN and/or uncertainty-aware training did not improve calibration and sometimes degraded performance, both in terms of accuracy and calibration.

For the PC-CMR experiment, the higher capacity ResNet18 model performed better than the low capacity LeNet5 model in terms of accuracy. For the ResNet18, the ENN alone and the ENN combined with UvAC uncertainty aware training provided a significant improvement to model calibration but the UvAC alone did not. In contrast, for the LeNet5, only the combination of the ENN and UvAC uncertainty-aware training improved calibration, whilst the ENN or UvAC alone did not. The only approach that improved calibration for both the LeNet5 and ResNet18 models was the combination of the ENN and UvAC uncertainty-aware training.

We speculate that model calibration can suffer when the task becomes more challenging, requiring a higher capacity model (as in the PC-CMR experiment, compared to the easier MNIST task). In such situations it can be beneficial to apply a method to improve model calibration. For example, noticeable improvements are evident in the reliability diagrams in Figure \ref{fig:f1} c-d over the baseline ResNet18 in Figure \ref{fig:f1}a. Based on our results it seems as if the best way to improve model calibration may be dependent on the capacity of the model, although the combination of ENN and uncertainty-aware training seems to be robust to differences in model capacity.

To the best of our knowledge the combination of an ENN with uncertainty-aware training has not previously been investigated in a medical imaging context but such a combination has been proposed in a computer vision context with some preliminary success \cite{bao2021evidential}. Our results indicate that a robust and deterministic evidence-based and uncertainty-aware learning approach may improve calibration of DL models by improving the modelling of uncertainty. In future work we will investigate other uncertainty-aware training approaches in order to develop trustworthy and reliable DL models for healthcare applications.

\begin{table*}[htb]
 \caption{Accuracy and calibration results of all models applied to the MNIST and PC-CMR datasets. Evaluation metrics are: test accuracy (Acc) and Adaptive Expected Calibration Error (AECE) in \textcolor{blue}{blue}. Noise level indicated is the standard deviation of the Gaussian distribution. All results averaged over 10 train/test runs.}
  \centering
  \small
  \begin{adjustbox}{max width=\textwidth}
  \begin{tabular}{l||c|c|c|c|c|c||c}
    \toprule
    \midrule
    &\multicolumn{5}{c}{\textbf{MNIST (Acc, \textcolor{blue}{AECE})}} 
    &\multicolumn{1}{c}{\textbf{PC-CMR (Acc, \textcolor{blue}{AECE})}}\\
    \textbf{Method/Noise levels:} & 0 & 0.2 & 0.3 & 0.5 & 6 & \\
        \midrule
        \textbf{LeNet5} & 98.9 $\pm$ \num{0.1}, \textcolor{blue}{0.003 $\pm$ \num{0.001}} & 98.7 $\pm$ \num{0.05}, \textcolor{blue}{0.003 $\pm$ \num{0.001}} & 98.7 $\pm$ \num{0.07}, \textcolor{blue}{0.003 $\pm$ \num{0.0005}} & 98.3 $\pm$ \num{0.12}, \textcolor{blue}{0.003 $\pm$ \num{0.001}} & 62.5 $\pm$ \num{0.13} , \textcolor{blue}{0.01 $\pm$ \num{0.005}} & 52.0 $\pm$ \num{4.6}, \textcolor{blue}{0.12 $\pm$ \num{0.02}}\\
        \textbf{LeNet5+UvAC} & 99.0 $\pm$ \num{0.048}, \textcolor{blue}{0.004 $\pm$ \num{0.001}} & 98.9 $\pm$ \num{0.011}, \textcolor{blue}{0.005 $\pm$ \num{0.001}} & 98.8 $\pm$ \num{0.11} , \textcolor{blue}{0.004 $\pm$ \num{0.001}} & 99.0 $\pm$ \num{0.10}, \textcolor{blue}{0.004 $\pm$ \num{0.001}} & 63.1 $\pm$ \num{0.01}, \textcolor{blue}{0.01 $\pm$ \num{0.002}} &  50.0 $\pm$ \num{7.6}, \textcolor{blue}{0.11 $\pm$ \num{0.05}}\\ 
        \textbf{LeNet5+ENN} & 98.0 $\pm$ \num{0.06}, \textcolor{blue}{0.05 $\pm$ \num{0.04}} & 86.70 $\pm$ \num{12.0}, \textcolor{blue}{0.12 $\pm$ \num{0.10}} & 85.30 $\pm$ \num{9.2}, \textcolor{blue}{0.14 $\pm$ \num{0.10}} & 89.30 $\pm$ \num{8.4} \textcolor{blue}{0.1 $\pm$ \num{0.08}} & 57.40 $\pm$ \num{2.8}, \textcolor{blue}{0.21 $\pm$ \num{0.03}} & 54.60 $\pm$ \num{7.0}, \textcolor{blue}{0.12 $\pm$ \num{0.05}}\\ 
        \textbf{LeNet5+ENN+UvAC} & 97.1 $\pm$ \num{2.90}, \textcolor{blue}{0.03 $\pm$ \num{0.03}} & 93.20 $\pm$ \num{4.37} \textcolor{blue}{0.06 $\pm$ \num{0.04}} & 82.04 $\pm$ \num{7.42}, \textcolor{blue}{0.17} $\pm$ \num{0.06} & 84.91 $\pm$ \num{9.47}, \textcolor{blue}{0.14 $\pm$ \num{0.10}} & 50.89 $\pm$ \num{6.55}, \textcolor{blue}{0.33 $\pm$ \num{0.04}} & 55.1 $\pm$ \num{8.9}, \textcolor{blue}{0.06 $\pm$ \num{0.03}}\\ 
        \midrule
        \textbf{ResNet18} & 99.0 $\pm$ \num{0.04}, \textcolor{blue}{0.003 $\pm$ \num{0.001}} & 98.9 $\pm$ \num{0.09}, \textcolor{blue}{0.003 $\pm$ \num{0.001}} & 98.9 $\pm$ \num{0.09}, \textcolor{blue}{0.003 $\pm$ \num{0.001}} & 98.8 $\pm$ \num{0.14}, \textcolor{blue}{0.002 $\pm$ \num{0.001}} & 61.6, $\pm$ \num{0.37}, \textcolor{blue}{0.01 $\pm$ \num{0.002}} & 64.8 $\pm$ \num{5}, \textcolor{blue}{0.17 $\pm$ \num{0.05}}\\ 
        \textbf{ResNet18+UvAC} & 99.0 $\pm$ \num{0.005}, \textcolor{blue}{0.003 $\pm$ \num{0.001}} & 99.0 $\pm$ \num{0.004}, \textcolor{blue}{0.003 $\pm$ \num{0.001}} & 99.0 $\pm$ \num{0.11}, \textcolor{blue}{0.003 $\pm$ \num{0.001}} & 99.0 $\pm$ \num{0.1}, \textcolor{blue}{0.002 $\pm$ \num{0.001}} & 62.0 $\pm$ \num{0.37}, \textcolor{blue}{0.01 $\pm$ \num{0.002}} & 64.5$\pm$ \num{3.2}, \textcolor{blue}{0.17 $\pm$ \num{0.04}}\\  
        \textbf{ResNet18+ENN} & 99.0 $\pm$ \num{0.35}, \textcolor{blue}{0.01 $\pm$ \num{0.01}} & 99.2 $\pm$ \num{0.08}, \textcolor{blue}{0.008 $\pm$ \num{0.002}} & 99.0 $\pm$ \num{0.09}, \textcolor{blue}{0.01 $\pm$ \num{0.01}} & 98.4 $\pm$ \num{1.58}, \textcolor{blue}{0.01 $\pm$ \num{0.01}} & 61.0 $\pm$ \num{3.13}, \textcolor{blue}{0.13 $\pm$ \num{0.01}} & 66.0 $\pm$ \num{4.90}, \textcolor{blue}{0.06 $\pm$ \num{0.02}} \\  
        \textbf{ResNet18+ENN+UvAC} & 98.7 $\pm$ \num{0.35}, \textcolor{blue}{0.05 $\pm$ \num{0.03}} & 98.8 $\pm$ \num{0.13}, \textcolor{blue}{0.05 $\pm$ \num{0.01}} & 99.0 $\pm$ \num{0.17}, \textcolor{blue}{0.06 $\pm$ \num{0.01}} & 98.7 $\pm$ \num{0.23}, \textcolor{blue}{0.07 $\pm$ \num{0.03}} & 61.4 $\pm$ \num{0.53}, \textcolor{blue}{0.31 $\pm$ \num{0.05}} & 66.0 $\pm$ \num{4.70}, \textcolor{blue}{0.06 $\pm$ \num{0.02}} \\ 
        \bottomrule
    \end{tabular}
    \label{table: Evaluation Results MNIST PC CMR Flow}
\end{adjustbox}
\end{table*}





\section{Compliance with ethical standards}
\label{sec:ethics}
This research study was conducted retrospectively using human subject data made available using the UK Biobank Resource under Application Number 17806. 





\section{Acknowledgments}
\label{sec:acknowledgments}

This work was supported by the Kings DRIVE Health CDT for Data-Driven Health and further funded/supported by the National Institute for Health Research (NIHR) Biomedical Research Centre at Guy’s and St Thomas’ NHS Foundation Trust and King’s College London. Additionally this research was funded in whole, or in part, by the Wellcome Trust WT203148/Z/16/Z. For the purpose of open access, the author has applied a CC BY public copyright license to any author accepted manuscript version arising from this submission. This research has been conducted using the UK Biobank Resource under Application Number 17806. 

The views expressed in this paper are those of the authors and not necessarily those of the NHS, EPSRC, the NIHR or the Department of Health and Social Care. 






\bibliographystyle{IEEEbib}
\bibliography{refs_original}

\begin{thebibliography}{10}

\bibitem{chen2022recent}
Xuxin Chen, Ximin Wang, Ke~Zhang, et~al.,
\newblock ``Recent advances and clinical applications of deep learning in
  medical image analysis,''
\newblock {\em Medical Image Analysis}, p. 102444, 2022.

\bibitem{jin2022guidelines}
Weina Jin, Xiaoxiao Li, Mostafa Fatehi, and Ghassan Hamarneh,
\newblock ``Guidelines and evaluation for clinical explainable ai on medical
  image analysis,''
\newblock {\em arXiv preprint arXiv:2202.10553}, 2022.

\bibitem{nixon2019measuring}
Jeremy Nixon, Michael~W Dusenberry, Linchuan Zhang, et~al.,
\newblock ``Measuring calibration in deep learning.,''
\newblock in {\em CVPR Workshops}, 2019, vol.~2.

\bibitem{reinke2023understanding}
Annika Reinke, Minu~D Tizabi, Michael Baumgartner, Matthias Eisenmann, Doreen
  Heckmann-N{\"o}tzel, A~Emre Kavu, Tim R{\"a}dsch, Carole~H Sudre, Laura
  Acion, Michela Antonelli, et~al.,
\newblock ``Understanding metric-related pitfalls in image analysis
  validation,''
\newblock {\em arXiv preprint arXiv:2302.01790}, 2023.

\bibitem{kompa2021second}
Benjamin Kompa, Jasper Snoek, and Andrew~L Beam,
\newblock ``Second opinion needed: communicating uncertainty in medical machine
  learning,''
\newblock {\em NPJ Digital Medicine}, vol. 4, no. 1, pp. 1--6, 2021.

\bibitem{gawlikowski2021survey}
Jakob Gawlikowski, Cedrique Rovile~Njieutcheu Tassi, et~al.,
\newblock ``A survey of uncertainty in deep neural networks,''
\newblock {\em arXiv preprint arXiv:2107.03342}, 2021.

\bibitem{sensoy2018evidential}
Murat Sensoy, Lance Kaplan, and Melih Kandemir,
\newblock ``Evidential deep learning to quantify classification uncertainty,''
\newblock {\em Advances in neural information processing systems}, vol. 31,
  2018.

\bibitem{guo2017calibration}
Chuan Guo, Geoff Pleiss, Yu~Sun, and Kilian~Q Weinberger,
\newblock ``On calibration of modern neural networks,''
\newblock in {\em International conference on machine learning}. PMLR, 2017,
  pp. 1321--1330.

\bibitem{bao2021evidential}
Wentao Bao, Qi~Yu, and Yu~Kong,
\newblock ``Evidential deep learning for open set action recognition,''
\newblock in {\em Proceedings of the IEEE/CVF International Conference on
  Computer Vision}, 2021, pp. 13349--13358.

\bibitem{minary2017face}
Pauline Minary, Fr{\'e}d{\'e}ric Pichon, et~al.,
\newblock ``Face pixel detection using evidential calibration and fusion,''
\newblock {\em International Journal of Approximate Reasoning}, vol. 91, pp.
  202--215, 2017.

\bibitem{tomani2021towards}
Christian Tomani and Florian Buettner,
\newblock ``Towards trustworthy predictions from deep neural networks with fast
  adversarial calibration,''
\newblock in {\em Proceedings of the AAAI Conference on Artificial
  Intelligence}, 2021, vol.~35, pp. 9886--9896.

\bibitem{karandikar2021soft}
Archit Karandikar, Nicholas Cain, et~al.,
\newblock ``Soft calibration objectives for neural networks,''
\newblock {\em Advances in Neural Information Processing Systems}, vol. 34,
  2021.

\bibitem{krishnan2020improving}
Ranganath Krishnan and Omesh Tickoo,
\newblock ``Improving model calibration with accuracy versus uncertainty
  optimization,''
\newblock {\em Advances in Neural Information Processing Systems}, 2020.

\bibitem{dawood2021uncertainty}
Tareen Dawood, Chen Chen, Robin Andlauer, et~al.,
\newblock ``Uncertainty-aware training for cardiac resynchronisation therapy
  response prediction,''
\newblock in {\em Statistical Atlases and Computational Models of the Heart}.
  2022, pp. 189--198, Springer International Publishing.

\bibitem{yu2019uncertainty}
Lequan Yu, Shujun Wang, Xiaomeng Li, Chi-Wing Fu, and Pheng-Ann Heng,
\newblock ``Uncertainty-aware self-ensembling model for semi-supervised {3D}
  left atrium segmentation,''
\newblock in {\em International Conference on Medical Image Computing and
  Computer-Assisted Intervention}. Springer, 2019, pp. 605--613.

\bibitem{sudlow2015uk}
Cathie Sudlow, John Gallacher, Naomi Allen, et~al.,
\newblock ``Uk biobank: an open access resource for identifying the causes of a
  wide range of complex diseases of middle and old age,''
\newblock {\em PLoS medicine}, vol. 12, no. 3, pp. e1001779, 2015.

\bibitem{https://doi.org/10.48550/arxiv.2209.14212}
Emily Chan, Ciaran O'Hanlon, Carlota~Asegurado Marquez, et~al.,
\newblock ``Automated quality controlled analysis of 2d phase contrast
  cardiovascular magnetic resonance imaging,''
\newblock in {\em Statistical Atlases and Computational Models of the Heart (In
  Press)}, 2022.

\bibitem{lecun1998gradient}
Yann LeCun, L{\'e}on Bottou, Yoshua Bengio, and Patrick Haffner,
\newblock ``Gradient-based learning applied to document recognition,''
\newblock {\em Proceedings of the IEEE}, vol. 86, no. 11, pp. 2278--2324, 1998.

\bibitem{he2016deep}
Kaiming He, Xiangyu Zhang, Shaoqing Ren, and Jian Sun,
\newblock ``Deep residual learning for image recognition,''
\newblock in {\em Proceedings of the IEEE conference on computer vision and
  pattern recognition}, 2016, pp. 770--778.

\bibitem{ding2020revisiting}
Yukun Ding, Jinglan Liu, Jinjun Xiong, and Yiyu Shi,
\newblock ``Revisiting the evaluation of uncertainty estimation and its
  application to explore model complexity-uncertainty trade-off,''
\newblock in {\em Proceedings of the IEEE/CVF Conference on Computer Vision and
  Pattern Recognition Workshops}, 2020, pp. 4--5.

\bibitem{ciresan2011flexible}
Dan~Claudiu Ciresan, Ueli Meier, Jonathan Masci, Luca~Maria Gambardella, and
  J{\"u}rgen Schmidhuber,
\newblock ``Flexible, high performance convolutional neural networks for image
  classification,''
\newblock in {\em Twenty-second international joint conference on artificial
  intelligence}, 2011.

\end{thebibliography}
\end{document}